\documentclass[10pt,journal]{IEEEtran}

\usepackage{amsmath}
\usepackage{amssymb}
\usepackage{bm}
\usepackage{color}
\usepackage{graphicx}
\usepackage{empheq}
\usepackage[linesnumbered,ruled,lined]{algorithm2e}
\usepackage[outdir=./]{epstopdf}
\usepackage{caption}
\usepackage{lipsum}
\usepackage{cite}
\usepackage{multirow}
\usepackage{hyperref}
\hypersetup{hidelinks,
	colorlinks=true,
	allcolors=black,
	pdfstartview=Fit,
	breaklinks=true}
\usepackage{subcaption}
\usepackage{fancyhdr}
\usepackage{graphicx}
\SetKwInput{Input}{input}
\SetKwInput{Output}{output}

\usepackage{array}
\newcolumntype{L}[1]{>{\raggedright\let\newline\\\arraybackslash\hspace{0pt}}m{#1}}
\newcolumntype{C}[1]{>{\centering\let\newline\\\arraybackslash\hspace{0pt}}m{#1}}
\newcolumntype{R}[1]{>{\raggedleft\let\newline\\\arraybackslash\hspace{0pt}}m{#1}}
\IEEEoverridecommandlockouts

\begin{document}

\title{Rate-Splitting Multiple Access with Finite Blocklength for Short-Packet and Low-Latency Downlink Communications}
\makeatletter
\def\ps@IEEEtitlepagestyle{
    \def\@oddfoot{\mycopyrightnotice}
    \def\@evenfoot{}
}
\def\mycopyrightnotice{
    {\footnotesize
            \begin{minipage}{\textwidth}
                \centering
                Copyright~\copyright~2015 IEEE. Personal use of this material is permitted. However, permission to use this \\
                material for any other purposes must be obtained from the IEEE by sending a request to pubs-permissions@ieee.org.
            \end{minipage}
        }
}
\author{

\IEEEauthorblockN{Yunnuo~Xu, Yijie~Mao, \IEEEmembership{Member, IEEE}, Onur~Dizdar, \IEEEmembership{Member, IEEE} and~Bruno~Clerckx, \IEEEmembership{Fellow, IEEE}\vspace{-1em}}

\thanks{(\textit{Corresponding author: Yijie Mao.})
\par Y. Xu, O. Dizdar and B. Clerckx are with Imperial College London, London SW7 2AZ, UK (email: {yunnuo.xu19, o.dizdar, b.clerckx}@imperial.ac.uk).
\par Y. Mao is with the School of Information Science and Technology, ShanghaiTech University, Shanghai 201210, China (email: maoyj@shanghaitech.edu.cn).

This work has been partially supported by the U.K. Engineering and Physical Sciences Research Council (EPSRC) under grant EP/N015312/1, EP/R511547/1.}
}
\maketitle

\begin{abstract}
    Rate-Splitting Multiple Access (RSMA) is an emerging flexible and powerful multiple access for downlink multi-antenna networks.
    In this paper, we introduce the concept of RSMA into short-packet  downlink communications. We design optimal linear precoders that maximize the sum rate with Finite Blocklength (FBL) constraints. The relations between the sum rate and blocklength of RSMA are investigated for a wide range of network loads and user deployments. Numerical results demonstrate that RSMA can achieve the same transmission rate as Non-Orthogonal Multiple Access (NOMA) and Space Division Multiple Access (SDMA) with smaller blocklengths (and therefore lower latency), especially in overloaded multi-antenna networks. Hence, we conclude that RSMA is a promising multiple access for  low-latency communications.
\end{abstract}
\vspace{-2mm}
\section{Introduction}
5G brings new features and enhancements compared to the previous generations of wireless communication networks, including higher data rates, Ultra-Reliable Low-Latency Communication (URLLC) and Machine Type Communication (MTC) \cite{Ultra-reliable_communication_in_5G_wireless_systems}. There are many latency-sensitive applications, such as industrial automation, smart grid and intelligent transportation.
In order to reduce the transmission latency, short-packets with Finite Blocklength (FBL) codes are typically adopted \cite{Toward_Massive_Ultrareliable_and_Low-Latency_Wireless_Communication_With_Short_Packets}. This brings a stringent latency requirement to the physical layer.
\par  In the pioneering work \cite{5452208}, Polyanskiy et al. provided information-theoretic limits on the achievable rate for given FBL and Block Error Rate (BLER).
Inspired by \cite{5452208}, the influence of FBL on the performance of various point-to-point communication scenarios has been investigated in \cite{7156144,6802432}. The maximum channel coding rate for both quasi-static fading channels and Additive White Gaussian Noise (AWGN) channels was studied in \cite{7156144} with constraints on BLER, blocklength and a long-term transmit power. In \cite{6802432}, the authors investigated the maximum achievable transmission rate over quasi-static Multiple-Input Multiple-Output (MIMO) fading channels under both perfect and imperfect Channel State Information at the Transmitter (CSIT)/Channel State Information at the Receiver (CSIR) settings.
\par Under the assumption of infinite blocklength, it is known that Non-Orthogonal Multiple Access (NOMA) utilizing Superposition Coding with Successive Interference Cancellation (SC-SIC) accommodates multiple users via non-orthogonal resource allocation \cite{7263349}. NOMA is well-known to outperform Orthogonal Multiple Access (OMA) with infinite blocklength \cite{1054727} and FBL \cite{8345745} communications. In multi-user multi-antenna systems, however, multi-antenna NOMA is shown to be an inefficient strategy in terms of multiplexing gain and use of SIC receivers \cite{clerckx2021noma}. Recently, Rate-Splitting Multiple Access (RSMA) has emerged as a promising multiple access for downlink multi-antenna networks under infinite blocklength \cite{7470942}, since it unifies and outperforms Space Division Multiple Access (SDMA), OMA, NOMA and physical-layer multicasting in multi-antenna networks for a wide range of network loads and user deployments \cite{8907421,Lina_Rate-Splitting_for_Downlink_Multi-User_Multi-Antenna_Systems}. Furthermore, RSMA provides significant gains in terms of spectral efficiency, energy efficiency, robustness and Channel State Information (CSI) feedback overhead reduction \cite{Lina_Rate-Splitting_for_Downlink_Multi-User_Multi-Antenna_Systems,7152864,7555358}, which are also important in the context of low-latency communications.
\par Motivated by the benefits of RSMA in the infinite blocklength regime and the significant performance gains of RSMA in link-level evaluations with practical codes \cite{dizdar2020rate}, we introduce RSMA into FBL downlink communications in this paper. All existing works of RSMA are based on the Shannon theorem with infinite blocklength. To the best of our knowledge, this is the first paper studying and optimizing RSMA for FBL communications. To that end, we leverage the information-theoretic limits provided in \cite{5452208}, and investigate the achievable sum rate under the constraints on blocklength and transmit power. Our results show that by utilizing RSMA for FBL downlink communications, the blocklength can be reduced, and the latency is therefore decreased, compared to SDMA and NOMA for the same transmission rate. Alternatively, for a given blocklength and therefore latency, RSMA can achieve a higher sum rate than SDMA and NOMA.

\vspace{-3.7mm}
\section{System model and problem formulation}\label{sec:System_model}
\vspace{-1.5mm}
\par We consider a system that consists of a Base Station (BS) equipped with $N_t$ antennas and $K$ single-antenna users. The users are indexed by $\mathcal{K}=\{1,2, \ldots,K\}$. The 1-layer RSMA scheme is adopted to serve $K$ users in this paper.\footnote{1-layer RSMA is denoted as RSMA in the rest of the paper unless otherwise stated. More complex Rate-Splitting (RS) architectures such as the generalized RS proposed in \cite{Lina_Rate-Splitting_for_Downlink_Multi-User_Multi-Antenna_Systems} can be considered for future works.} Assume that the BS intends to transmit $K$ messages to $K$ users. The message intended for user-$k$ is denoted by $W_{k}$, $\forall k \in \mathcal{K}$. The $W_{k}$ is divided into two parts, $\{W_{c,k},W_{p,k}\}$, which are named as the common part and the private part, respectively. Messages $\{W_{c,1}, \ldots, W_{c,K}\}$ are encoded together into the common stream $s_{c}$ to be decoded by all $K$ users. The private message $W_{p,k}$ is encoded into the private stream $s_k$. The overall symbol stream vector to be transmitted is denoted by $\mathbf{s}={[s_{c},s_1,s_2, \ldots,s_K]}^T$. The symbols are precoded via a precoding matrix $\mathbf{P}=[\mathbf{p}_{c},\mathbf{p}_1,\mathbf{p}_2, \ldots,\mathbf{p}_K]$, where $\mathbf{p}_j\in\mathbb{C}^{N_t\times1}$ represents the linear precoder for the stream $s_j$, $j\in\{c,1,2, \ldots,K\}$. The transmit signal can be expressed by
\begin{equation}
    \setlength{\abovedisplayskip}{0.5pt}
    \setlength{\belowdisplayskip}{0.5pt}
    \mathbf{x}=\mathbf{Ps}=\mathbf{p}_{c}s_c+\sum\nolimits_{k\in\mathcal{K}}\mathbf{p}_ks_k.
\end{equation}
We assume that the symbols have unit power, i.e., $\mathbb{E}(\mathbf{s}\mathbf{s}^H)=\mathbf{I}$. The transmit signal is subject to a maximum transmit power $P_t$, such that, $\mathbb{E}\{\mathbf{x}\mathbf{x}^H\}\leq P_t$, and equivalently $\mathrm{tr}(\mathbf{P}\mathbf{P}^H)\leq P_t$. The received signal $y_k$ at user-\textit{k} is written as
\begin{equation}
    \setlength{\abovedisplayskip}{0.5pt}
    \setlength{\belowdisplayskip}{0.5pt}
    y_k=\mathbf{h}_k^H \mathbf{x}+n_k,
\end{equation}
where $\mathbf{h}_k\in\mathbb{C}^{N_t\times1}$ is the channel between the BS and user-\textit{k}, and term $n_k\sim\mathcal{CN}(0,1)$ is AWGN at user-$k$. We assume perfect CSIT and perfect CSIR.
\par Each user first decodes the common stream $s_{c}$ by treating the interference from the private streams $s_1, \ldots,s_K$ as noise. Accordingly, the Signal to Interference plus Noise Ratio (SINR) of the common stream at user-$k$ is
\begin{equation}
    \setlength{\abovedisplayskip}{0.5pt}
    \setlength{\belowdisplayskip}{0.5pt}
    \Gamma_{c,k}=\frac{{|\mathbf{h}_k^H\mathbf{p}_{c}|}^2}{\sum_{j\in\mathcal{K}}{|\mathbf{h}_k^H\mathbf{p}_j|}^2+1}.
\end{equation}
\par Assuming the common stream $s_c$ is successfully decoded, it is reconstructed and removed from the received signal $y_k$. Subsequently, user-$k$ decodes the private stream $s_k$ while treating the private streams of other users as noise. Accordingly, the SINR of decoding the private stream $s_k$ at user-$k$ is
\begin{equation}
    \setlength{\abovedisplayskip}{0.5pt}
    \setlength{\belowdisplayskip}{0.5pt}
    \Gamma_{p,k}=\frac{{|\mathbf{h}_k^H\mathbf{p}_k|}^2}{\sum_{j\in\mathcal{K},j\neq k}{|\mathbf{h}_k^H\mathbf{p}_j|}^2+1}.
\end{equation}
\par Once the common and private streams are decoded, user-$k$ reconstructs the original message by extracting the decoded $W_{c,k}$ from the decoded $W_{c}$, and combining the decoded $W_{c,k}$ with the decoded $W_{p,k}$. Following \cite{5452208}, the respective rates $R_{c,k}, R_{p,k}$ of the common and private streams are
\begin{subequations}\label{Equ:total_Rate}
    \setlength{\abovedisplayskip}{0.5pt}
    \setlength{\belowdisplayskip}{0.5pt}
    \begin{align}
         & R_{c,k} \approx \log_2(1+\Gamma_{c,k})-(\log_2e)\sqrt{\frac{V_{c,k}}{l_{c}}}Q^{-1}(\epsilon)\label{total_Rate_common},  \\
         & R_{p,k} \approx \log_2(1+\Gamma_{p,k})-(\log_2e)\sqrt{\frac{V_{p,k}}{l_{k}}}Q^{-1}(\epsilon)\label{total_Rate_private},
    \end{align}
\end{subequations}
where $l_c,l_1,\ldots,l_K$ are the respective blocklengths of streams $s_c,s_1,\ldots,s_K$ and $\epsilon$ represents the BLER. The function $Q^{-1}(.)$ corresponds to the inverse of the Gaussian $\mathcal{Q}$ function\footnote{$Q(x)=\int_x^{\infty}\frac{1}{\sqrt{2\pi}}\exp(-\frac{t^2}{2})dt$.}. $V_{c,k}$ and $V_{p,k}$ are the channel dispersion parameters with expressions given by
\begin{equation}\label{Equ:ChannelDisper}
    \setlength{\abovedisplayskip}{0.2pt}
    \setlength{\belowdisplayskip}{0.2pt}
    V_{i,k} = 1-{(1+\Gamma_{i,k})}^{-2},\,i\in\{c,p\}.
\end{equation}
\par In this work, we consider signaling by Gaussian codebooks, so that the expression (\ref{Equ:total_Rate}) is valid for our setup. To ensure that the common stream is successfully decoded by all users, the achievable rate of the common stream (also denoted as common rate) shall not exceed $R_{c}=\min\{R_{c,1},R_{c,2}, \ldots,R_{c,K}\}$.
\par We focus on the precoder design for FBL signal $\mathbf{x}$ to maximize the sum rate subject to blocklength and transmit power constraints. The blocklength for each stream is $l$, i.e., $l_j= l,\,j\in\{c,1,2, \ldots,K\}$. The blocklength of the codeword is proportional to the latency and can be approximately expressed as, $l\approx BT$, where $B$ and $T$ are the bandwidth and time duration of the signal (i.e., latency), respectively \cite{Toward_Massive_Ultrareliable_and_Low-Latency_Wireless_Communication_With_Short_Packets}. In this work, we fix the bandwidth $B$ and investigate the effect of blocklength. For a given blocklength $l$, the achievable sum rate of $K$-user RSMA is
\begin{subequations}\label{Prob:k_user_example}
    \setlength{\abovedisplayskip}{0.5pt}
    \setlength{\belowdisplayskip}{0.5pt}
    \begin{align}
        \max_{\mathbf{P},\mathbf{c}} \quad & \sum_{k\in\mathcal{K}}R_{k,tot}                                                                              \\
        \mbox{s.t.}\quad
                                           & \sum_{k^{\prime}\in\mathcal{K}}C_{k^{\prime}}\leq R_{c,k},\,\forall k\in\mathcal{K}\label{k_user_example_c1} \\
                                           & \mathrm{tr}(\mathbf{P}\mathbf{P}^{H})\leq P_t\label{k_user_example_c2}                                       \\
                                           & R_{k,tot}\geq r_{k}^{th},\,\forall k\in\mathcal{K}\label{QoS}                                                \\
                                           & \mathbf{c}\geq\mathbf{0}\label{k_user_example_cc},
    \end{align}
\end{subequations}
where $\mathbf{c}=[C_1,C_2, \ldots,C_K]$ and $C_k$ is the \textit{k}th user's portion of the common rate with $\sum_{k\in\mathcal{K}}C_k=R_c$. The total transmission rate of user-\textit{k} is $R_{k,tot}=C_k+R_{p,k}$. The constraint (\ref{QoS}) guarantees the Quality-of-Service (QoS) for the users by the individual rate constraint $r_{k}^{th}$, $\forall k \in \mathcal{K}$. The sum rate maximization problem (\ref{Prob:k_user_example}) is a non-convex fractional program. Considering the asymptotic case when $l$ goes to infinity, the second terms in $R_{c,k}$ and $R_{p,k}$ become $0$ and the optimization problem (\ref{Prob:k_user_example}) is simplified to the one considered in \cite{Lina_Rate-Splitting_for_Downlink_Multi-User_Multi-Antenna_Systems} with infinite blocklength. Therefore, the optimization problem (\ref{Prob:k_user_example}) is more general than the one investigated in \cite{Lina_Rate-Splitting_for_Downlink_Multi-User_Multi-Antenna_Systems}. By comparing sum rate expressions of FBL and infinite blocklength scenarios, it is observed that the sum rate loss results from second terms in the common and private rate.

\vspace{-6mm}
\section{Proposed algorithm}\label{sec:proposed_algorithm}
\vspace{-3mm}
\par In this section, we describe the Successive Convex Approximation (SCA) algorithm adopted to jointly optimize precoders $\mathbf{P}$ and the common rate vector $\mathbf{c}$. SCA has been introduced in \cite{9123680} for RSMA precoder design in cooperative systems. The proposed SCA approach differs from the existing one in \cite{9123680} due to the FBL rate expressions in (\ref{Equ:total_Rate}) and the non-convex nature of the channel dispersion expression, which brings further challenge to approximation.
\par The problem (\ref{Prob:k_user_example}) is non-convex due to the non-convex rate expressions. We introduce variables $\boldsymbol{\beta}_p=[\beta_{p,1},\allowbreak\beta_{p,2},\allowbreak\ldots,\allowbreak\beta_{p,K}]$, $\boldsymbol{\rho}_c=[\rho_{c,1},\allowbreak\rho_{c,2},\allowbreak\ldots,\allowbreak\rho_{c,K}]$, $\boldsymbol{\rho}_p=[\rho_{p,1},\allowbreak\rho_{p,2}, \allowbreak\ldots,\allowbreak\rho_{p,K}]$, $\boldsymbol{\sigma}_{c}=[\sigma_{c,1},\allowbreak\sigma_{c,2},\allowbreak\ldots,\allowbreak\sigma_{c,K}]$ and $\boldsymbol{\sigma}_{p}=[\sigma_{p,1},\allowbreak\sigma_{p,2},\allowbreak\ldots,\allowbreak\sigma_{p,K}]$, where $\beta_{p,k}$ is the lower bound of the private rate $R_{p,k}$, $\rho_{c,k}$ and $\rho_{p,k}$ are the lower bounds of the SINR of the common and private streams, respectively, and $\sigma_{c,k}$ and $\sigma_{p,k}$ are the upper bounds for the interference plus noise terms corresponding to $\rho_{c,k}$ and $\rho_{p,k}$, respectively. For a given BLER and blocklength, $D=\frac{{Q}^{-1}(\epsilon)}{\sqrt{l}}\log_2e$ is a constant. We also introduce the dispersion parameter in terms of $\rho_{i,k}$ as $\nu_{i,k}=1-(1+\rho_{i,k})^{-2},\,i\in\{c,p\}$. With the introduced variables, Problem (\ref{Prob:k_user_example}) is equivalently written as (\ref{Prob:2nd_transform}).
\par Problem (\ref{Prob:2nd_transform}) remains non-convex due to the non-convex constraints (\ref{2nd_transform_c3})\verb|-|(\ref{2nd_transform_c6}). Next, we approximate the non-convex parts $\sqrt{\nu_{c,k}}$ and $\sqrt{\nu_{p,k}}$ in the constraints by the first-order Taylor series. Constraints (\ref{2nd_transform_c3}) and (\ref{2nd_transform_c4}) are approximated at $\rho_{c,k}^{[n]}$ and $\rho_{p,k}^{[n]}$ at iteration \textit{n} as (\ref{equ:taylor_rate}).
Constraints (\ref{2nd_transform_c5}) and (\ref{2nd_transform_c6}) are approximated at the points $(\mathbf{p}_{c}^{[n]},\sigma_{c,k}^{[n]})$ and $(\mathbf{p}_{k}^{[n]},\sigma_{p,k}^{[n]})$ respectively as (\ref{equ:sigma_expan}).
\begin{subequations}\label{Prob:2nd_transform}
    \setlength{\abovedisplayskip}{1pt}
    \setlength{\belowdisplayskip}{0.025pt}
    \begin{align}
        \max_{\substack{\mathbf{P},\mathbf{c},\boldsymbol{\beta}_p,\boldsymbol{\rho}_c,                                                                                                                             \\
        \boldsymbol{\rho}_p,\boldsymbol{\sigma}_{c},\boldsymbol{\sigma}_{p}}} & \quad\quad\sum_{k\in\mathcal{K}}(C_k+\beta_{p,k})\label{2nd_transform_c1}                                                           \\
        \mbox{s.t.}\quad\quad
        \begin{split}
            &\log_2(1+\rho_{c,k}) - D\sqrt{\nu_{c,k}} \geq\sum_{k^{\prime}\in\mathcal{K}}C_{k^{\prime}},\,\forall k\in\mathcal{K}
        \end{split}\label{2nd_transform_c3}                                            \\
        \begin{split}
            &\log_2(1+\rho_{p,k}) - D\sqrt{\nu_{p,k}} \geq \beta_{p,k},\,\forall k\in\mathcal{K}
        \end{split}\label{2nd_transform_c4}                                                                                                            \\
                                                                              & \frac{{|\mathbf{h}_k^H\mathbf{p}_{c}|}^2}{\sigma_{c,k}}\geq\rho_{c,k},\,\forall k\in\mathcal{K}\label{2nd_transform_c5}             \\
                                                                              & \frac{{|\mathbf{h}_k^H\mathbf{p}_{k}|}^2}{\sigma_{p,k}}\geq\rho_{p,k},\,\forall k\in\mathcal{K}\label{2nd_transform_c6}             \\
                                                                              & \sigma_{c,k}\geq\sum\nolimits_{j\in\mathcal{K}}{|\mathbf{h}_k^H\mathbf{p}_j|}^2+1,\,\forall k\in\mathcal{K}\label{2nd_transform_c7} \\
                                                                              & \sigma_{p,k}\geq\sum\nolimits_{\substack{j\in\mathcal{K},                                                                           \\j\neq k}}{|\mathbf{h}_k^H\mathbf{p}_j|}^2+1,\,\forall k\in\mathcal{K}\label{2nd_transform_c8}\\
                                                                              & C_k+\beta_{p,k}\geq r_{k}^{th},\,\forall k\in\mathcal{K}\label{QoS_2}                                                               \\
                                                                              & (\ref{k_user_example_c2}), (\ref{k_user_example_cc}).
        \vspace{-4mm}
    \end{align}
\end{subequations}
\begin{subequations}\label{equ:taylor_rate}
    \setlength{\abovedisplayskip}{0.025pt}
    \setlength{\belowdisplayskip}{0.05pt}
    \begin{align}
        \begin{split}
            &\log_2(1+\rho_{c,k})-D\bigg\{\left[1-(1+\rho^{[n]}_{c,k})^{{-2}}\right]^{-\frac{1}{2}}\Big[(1+\rho^{[n]}_{c,k})^{-3}\\
                &\,\quad\quad(\rho_{c,k}-\rho^{[n]}_{c,k})-(1+\rho^{[n]}_{c,k})^{-2}+1\Big]\bigg\}\geq \sum\nolimits_{k\in\mathcal{K}}C_k,
        \end{split} \\
        \begin{split}
            &\log_2(1+\rho_{p,k})-D\bigg\{\left[1-(1+\rho^{[n]}_{p,k})^{{-2}}\right]^{-\frac{1}{2}}\Big[(1+\rho^{[n]}_{p,k})^{-3}\\
                &\,\quad\quad(\rho_{p,k}-\rho^{[n]}_{p,k})-(1+\rho^{[n]}_{p,k})^{-2}+1\Big]\bigg\}\geq \beta_{p,k}.
        \end{split}
    \end{align}
\end{subequations}
\begin{subequations}\label{equ:sigma_expan}
    \setlength{\abovedisplayskip}{0.05pt}
    \setlength{\belowdisplayskip}{0.5pt}
    \begin{align}
         & \frac{2\mathfrak{R}\{(\mathbf{p}_c^{[n]})^H\mathbf{h}_k\mathbf{h}_k^H\mathbf{p}_c\}}{\sigma_{c,k}^{[n]}}-\frac{{|\mathbf{h}_k^H\mathbf{p}_c^{[n]}|}^2\sigma_{c,k}}{(\sigma_{c,k}^{[n]})^2}\geq\rho_{c,k}, \\
         & \frac{2\mathfrak{R}\{(\mathbf{p}_k^{[n]})^H\mathbf{h}_k\mathbf{h}_k^H\mathbf{p}_k\}}{\sigma_{p,k}^{[n]}}-\frac{{|\mathbf{h}_k^H\mathbf{p}_k^{[n]}|}^2\sigma_{p,k}}{(\sigma_{p,k}^{[n]})^2}\geq\rho_{p,k}.
    \end{align}
\end{subequations}
Based on the approximation methods described above, the original non-convex problem is transformed to a convex problem and can be solved using the SCA method. The main idea of SCA is to solve the non-convex problem by approximating it to a sequence of convex subproblems, which are solved successively. At iteration \textit{n}, based on the optimal solution ($\mathbf{P}^{[n-1]}$, $\boldsymbol{\rho}_c^{[n-1]}$, $\boldsymbol{\rho}_p^{[n-1]}$, $\boldsymbol{\sigma}_c^{[n-1]}$, $\boldsymbol{\sigma}_p^{[n-1]}$) obtained from the previous iteration $n-1$, we solve the following subproblem
\begin{subequations}\label{Prob:3rd_transform_final}
    \vspace{-4mm}
    \setlength{\abovedisplayskip}{0.5pt}
    \setlength{\belowdisplayskip}{0.5pt}
    \begin{align}
        \max_{\substack{\mathbf{P},\mathbf{c},\boldsymbol{\beta}_p,\boldsymbol{\rho}_c,                                       \\
        \boldsymbol{\rho}_p,\boldsymbol{\sigma}_{c},\boldsymbol{\sigma}_{p}}} \quad & \sum_{k\in\mathcal{K}}(C_k+\beta_{p,k}) \\
        \mbox{s.t.}\quad
        \begin{split}
            &(\ref{k_user_example_c2}),(\ref{k_user_example_cc}),(\ref{2nd_transform_c7}),\\
            &(\ref{2nd_transform_c8}),(\ref{QoS_2}),(\ref{equ:taylor_rate}), (\ref{equ:sigma_expan}).
        \end{split}
    \end{align}
\end{subequations}
Define $t=\sum_{k\in\mathcal{K}}(C_k+\beta_{p,k})$ for convenience, and $t^{[n]}=\sum_{k\in\mathcal{K}}(C_k^{[n]}+\beta_{p,k}^{[n]})$.
The proposed SCA-based algorithm is summarized in Algorithm \ref{Algor:WSR}. $\tau$ is the tolerance of algorithm. The convergence of Algorithm \ref{Algor:WSR} is guaranteed since the solution of Problem (\ref{Prob:3rd_transform_final}) at iteration $n-1$ is a feasible solution of the problem at iteration $n$. Therefore, the objective function $\sum_{k\in\mathcal{K}}(C_k+\beta_{p,k})$ is monotonically increasing and it is bounded above by the transmit power constraint.
\begin{algorithm}[t!]
    \setlength{\abovedisplayskip}{0.2pt}
    \setlength{\belowdisplayskip}{0.2pt}
    \caption{SCA-based algorithm}\label{Algor:WSR}
    \textbf{Initialize}: $n\leftarrow0$, $t^{[n]}\leftarrow0$, $\mathbf{P}^{[n]}$, $\boldsymbol{\rho}_c^{[n]}$, $\boldsymbol{\rho}_p^{[n]}$, $\boldsymbol{\sigma}_c^{[n]}$, $\boldsymbol{\sigma}_p^{[n]}$\;
    \Repeat{$|t^{[n]}-t^{[n-1]}|<\tau$}{
    $n\leftarrow n+1$\;
    Solve problem (\ref{Prob:3rd_transform_final}) using $\mathbf{P}^{[n-1]}$, $\boldsymbol{\rho}_c^{[n-1]}$, $\boldsymbol{\rho}_p^{[n-1]}$, $\boldsymbol{\sigma}_c^{[n-1]}$, $\boldsymbol{\sigma}_p^{[n-1]}$ and denote the optimal value of the objective function as ${t^{*}}$ and the optimal solutions as $\mathbf{P}^{*}$, $\boldsymbol{\rho}_c^{*}$, $\boldsymbol{\rho}_p^{*}$, $\boldsymbol{\sigma}_c^{*}$, $\boldsymbol{\sigma}_p^{*}$\;
    Update $t^{[n]}\leftarrow {t^{*}}$, $\mathbf{P}^{[n]}\leftarrow\mathbf{P}^{*}$, $\boldsymbol{\rho}_c^{[n]}\leftarrow\boldsymbol{\rho}_c^{*}$, $\boldsymbol{\rho}_p^{[n]}\leftarrow\boldsymbol{\rho}_p^{*}$, $\boldsymbol{\sigma}_c^{[n]}\leftarrow\boldsymbol{\sigma}_c^{*}$, $\boldsymbol{\sigma}_p^{[n]}\leftarrow\boldsymbol{\sigma}_p^{*}$\;
    }
\end{algorithm}
\par From (\ref{equ:taylor_rate}), it is noticed that the SINRs of the streams cannot be $0$ since the denominator of $[1-(1+\rho^{[n]}_{i,k})^{{-2}}]^{-\frac{1}{2}}$,$i\in\{c,p\}$ cannot be zero.  In a 2-user deployment, zero SINR corresponds to the scenarios when RSMA reduces to SDMA (i.e., the SINR of the common stream is $0$) or NOMA (i.e., the SINR of one of the private streams is $0$) \cite{8907421}.
In other words, when $K=2$ the proposed algorithm excludes the special cases when RSMA reduces to SDMA or NOMA. Therefore, Algorithm \ref{Algor:WSR} is denoted as an ``incomplete RSMA" algorithm. To deal with this problem, apart from the incomplete RSMA, we concurrently obtain the precoders from the specific regimes of RSMA by forcing the power allocated to one of streams to $0$, which corresponds to SDMA (by turning off the common stream) and NOMA (by turning off one private stream and allocating the entire common stream to transmit the message for a single user) in the 2-user scenario\footnote{When $K\geq3$, 1-layer RSMA can only boil down to SDMA. We obtain precoders from SDMA and incomplete RSMA.}. Finally, we select the scheme that achieves the highest sum rate among SDMA, NOMA and incomplete RSMA.

\vspace{-3.5mm}
\section{Results and discussion}\label{Sec:Results_and_discussion}
\vspace{-1.5mm}
The performances of RSMA, NOMA and SDMA with FBL in underloaded and overloaded scenarios are evaluated below.

\vspace{-5mm}
\subsection{Underloaded scenario}\label{Sec:Weighted sum rate}
\vspace{-1.5mm}
\subsubsection{Rate comparison}
In a two-user scenario, we first consider a simplified channel model of Uniform Linear Array (ULA) in order to investigate the effect of user angle and channel strength on the performance \cite{Lina_Rate-Splitting_for_Downlink_Multi-User_Multi-Antenna_Systems}.  When $N_t=4$, the channels between the BS and users are realized as $\mathbf{h}_1 = {[1,1,1,1]}^H, \mathbf{h}_2 = {\gamma\times[1,e^{j\theta},e^{j2\theta},e^{j3\theta}]}^H$, where $\gamma$ is the channel strength of user-2, and $\theta$ represents the difference in angle of departure between users, $\theta\in[0,\frac{\pi}{2}]$. For different $\gamma$, we investigate the channel realizations when $\theta\in\{\frac{\pi}{9},\frac{2\pi}{9},\frac{\pi}{3},\frac{4\pi}{9}\}$. SNR is fixed at 20 dB.
The sum rate and power allocation are investigated when the individual rate constraint changes as $\mathbf{r}_k^{th}=[0.01,\allowbreak0.1,\allowbreak0.15,\allowbreak0.17,\allowbreak0.19,\ldots,\allowbreak0.23,\allowbreak0.24,\allowbreak0.245,\allowbreak0.25,\allowbreak\ldots,\allowbreak0.27,\allowbreak0.273,\allowbreak0.276,\allowbreak\ldots,\allowbreak0.297,\allowbreak0.298,\allowbreak0.3]$ bit/s/Hz for $l=[100,200,300,\ldots,2500]$, $\forall k\in\{1,2\}$. It is known that Dirty Paper Coding (DPC) achieves the sum rate capacity of the Gaussian Broadcast Channel (BC) with infinite blocklength. In the following results, DPC is included as an upperbound of the sum rate, which is generated using the algorithm specified in \cite{9158344}.
Since RSMA and NOMA employ interference cancellation at user side, which may cause error propagation, we set the BLERs of RSMA and NOMA to $\epsilon_{RSMA}=\epsilon_{NOMA}=5\times 10^{-6}$ so as to guarantee the approximated overall BLER is not larger than $10^{-5}$. The BLER of SDMA is $\epsilon_{SDMA}=10^{-5}$.
\par
\begin{figure}[t!]
    \centering
    \begin{subfigure}[b]{0.48\linewidth}
        \hspace*{-0.37cm}\includegraphics[scale=0.6]{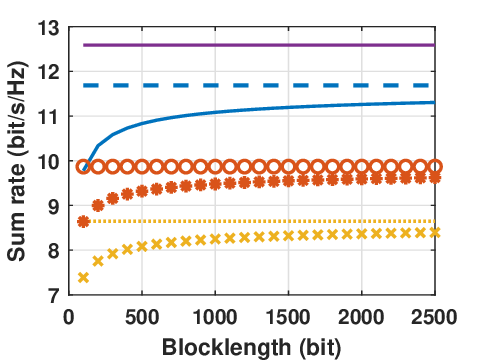}
        \caption{$\theta=\frac{\pi}{9}$}
    \end{subfigure}
    \begin{subfigure}[b]{0.48\linewidth}
        \hspace*{-0.1cm}\includegraphics[scale=0.6]{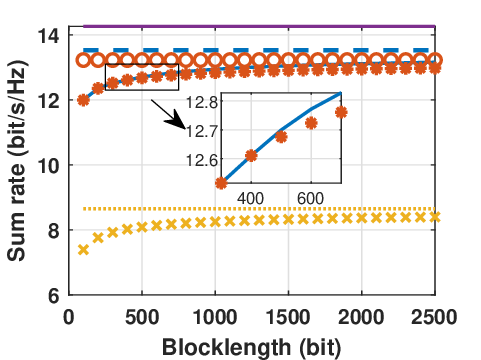}
        \caption{$\theta=\frac{2\pi}{9}$}
        \label{fig:sum_rate_bias1_2}
    \end{subfigure}
    \begin{subfigure}[b]{0.48\linewidth}
        \hspace*{-0.37cm}\includegraphics[scale=0.6]{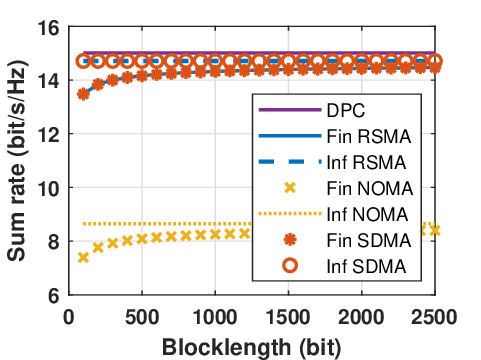}
        \caption{$\theta=\frac{3\pi}{9}$}
    \end{subfigure}
    \begin{subfigure}[b]{0.48\linewidth}
        \hspace*{-0.1cm}\includegraphics[scale=0.6]{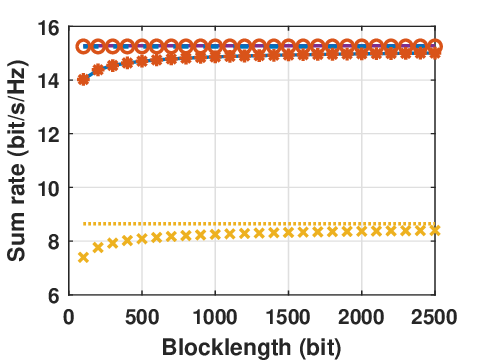}
        \caption{$\theta=\frac{4\pi}{9}$}
    \end{subfigure}
    \caption{Sum rate versus blocklength of different strategies for underloaded two-user deployment, $\gamma=1$.}
    \label{fig:sum_rate_bias1}
    \vspace{-5mm}
\end{figure}

 The notations ``Inf" and ``Fin" in figures represent the schemes when $l=\infty$ and when $l$ is finite, respectively. In Fig. \ref{fig:sum_rate_bias1}, the sum rate increases with the blocklength for all strategies. As the blocklength increases, the sum rate of all strategies approach that with infinite blocklength, as expected. As shown in Fig. \ref{fig:sum_rate_bias1}, RSMA outperforms NOMA and SDMA in all cases, especially when $\theta=\frac{\pi}{9}$. For the channel angles $\theta\in\{\frac{2\pi}{9},\frac{\pi}{3},\frac{4\pi}{9}\}$, the performance of RSMA is almost identical with SDMA, and both achieve better sum rate performance than NOMA. Furthermore, when $\theta=\frac{\pi}{9}$, RSMA requires a blocklength of 100 to achieve a sum rate of 9.7 bit/s/Hz, while SDMA needs a significantly longer blocklength of around 2500 to achieve the same sum rate. The results show that RSMA brings great benefits in terms of blocklength reduction as the user channels get more aligned with each other.
\par
\begin{figure}[t!]
    \centering
    \begin{subfigure}[b]{0.48\linewidth}
        \hspace*{-0.3cm}\includegraphics[scale=0.6]{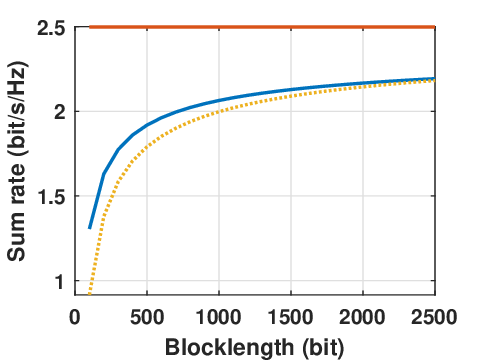}
        \caption{$\text{SNR}=0\,\text{dB}$}
    \end{subfigure}
    \begin{subfigure}[b]{0.48\linewidth}
        \hspace*{-0.1cm}\includegraphics[scale=0.6]{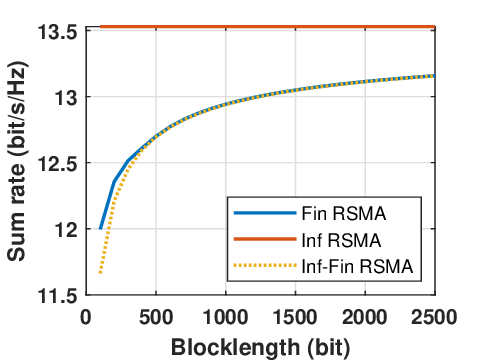}
        \caption{$\text{SNR}=20\,\text{dB}$}
        \label{fig:applying_inf_precoder_2}
    \end{subfigure}
    \caption{Applying the precoder from infinite blocklength scenario to FBL scheme with different SNR in underloaded two-user deployment, $\theta=\frac{2\pi}{9}$.}
    \label{fig:applying_inf_precoder}
    \vspace{-8mm}
\end{figure}

 We analyse the rate achieved by applying the precoder obtained under the infinite blocklength assumption to the scenario with FBL, i.e., ``Inf-Fin RSMA". From Fig. \ref{fig:applying_inf_precoder}, the gain of ``Fin RSMA" over ``Inf-Fin RSMA" is observed especially when the blocklength is small, and it is more obvious with small SNR value.

\subsubsection{Power allocation}
\par
\begin{figure}[t!]
    \centering
    \begin{subfigure}[b]{0.48\linewidth}
        \hspace*{-0.3cm}\includegraphics[scale=0.6]{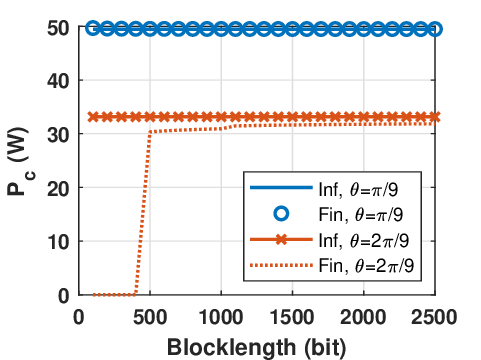}
        \caption{Power allocated to the\\\quad common stream}
        \label{fig:power_allo_merge_Pc}
    \end{subfigure}
    \begin{subfigure}[b]{0.48\linewidth}
        \hspace*{-0.1cm}\includegraphics[scale=0.6]{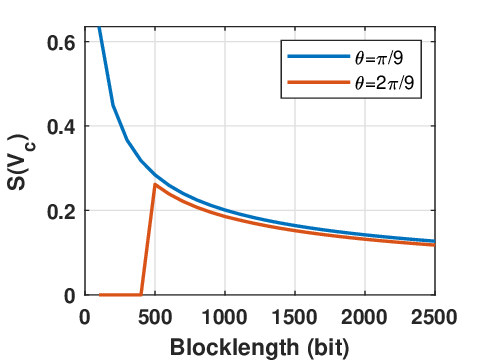}
        \caption{The second term $S(V_c)$ of the common stream}
        \label{fig:rate_loss}
    \end{subfigure}
    \caption{Power and $S(V_c)$ of the common message when $\theta\in\{\frac{\pi}{9},\frac{2\pi}{9}\}$, $\gamma=1$.}
    \vspace{-8mm}
\end{figure}

 The relation between power allocation of RSMA and blocklength is investigated in Fig. \ref{fig:power_allo_merge_Pc} for $\theta \in \{\frac{\pi}{9},\frac{2\pi}{9}\}$. An important observation from Fig. \ref{fig:power_allo_merge_Pc} is that the power allocated to the common stream tends to increase with the blocklength. Such phenomenon is explained as follows. When FBL is considered, the second term $S(x)=(\log_2e)\sqrt{\frac{x}{l}}Q^{-1}(\epsilon_{RSMA})$ (also known as the channel dispersion term) in (\ref{Equ:total_Rate}) reduces the achievable rate. Compared to SDMA, the additional stream introduced by RS (i.e., common stream) calls for power re-allocation (i.e., a part of $P_t$ may be allocated to the common stream). The use of common stream results in an additional second term $S(V_{c,k})$ in the sum rate expression, as evident from (\ref{total_Rate_common}). The splitting ratio is adjusted to compensate the term $S(V_{c,k})$ by decreasing the power allocated to the common stream (due to the monotonic decreasing behavior of the channel dispersion term with decreasing SINR). When the blocklength is reduced below a certain value, the power allocated to the common stream is adjusted and RS is not performed (i.e., zero power is allocated to the common stream), as observed in the case for $\theta=\frac{2\pi}{9}$ with a blocklength of 500 bits. For a blocklength that is smaller than 500 bits, the sum rate of SDMA is higher than that obtained by forcing to perform RS. This is visualized in Fig. \ref{fig:rate_loss}, which shows the term $S(V_{c})$ for the common stream, where $V_c=\arg\min_{V_{c,k}} R_{c,k}$. We first consider the case with $\theta=\frac{2\pi}{9}$, where RSMA achieves a sum rate gain of approximately 0.3 bit/s/Hz under infinite blocklength assumption, as observed from Fig. \ref{fig:sum_rate_bias1_2}. When $l\geq500$ and $\theta=\frac{2\pi}{9}$, $S(V_c)$ increases as the blocklength decreases, while in the range of $l<500$, RS is not performed due to the term $S(V_c)$ increasing beyond 0.3 bit/s/Hz. Consequently, the power allocated to the common stream is used for the private streams to achieve a better sum rate performance. The observation from Fig. \ref{fig:rate_loss} aligns with that from Fig. \ref{fig:power_allo_merge_Pc} and \ref{fig:sum_rate_bias1_2}.
On the other hand, the power allocated to the common stream does not change significantly with varying blocklength for $\theta=\frac{\pi}{9}$, as RSMA achieves a significant gain over SDMA and NOMA for the considered $\theta$ value with finite and infinite blocklength. When $\theta\geq\frac{3\pi}{9}$, RSMA does not achieve any sum rate gain over SDMA with infinite blocklength coding, thus RS is not performed for FBL coding, either.

\vspace{-6mm}
\subsection{Overloaded scenario}
\vspace{-1.4mm}
We investigate an overloaded scenario in this section. The performances of RSMA and SDMA are studied for special channel realizations and random channel realizations.
\par
\begin{figure}[t!]
    \centering
    \begin{subfigure}[b]{0.48\linewidth}
        \hspace*{-0.5cm}\includegraphics[scale=0.6]{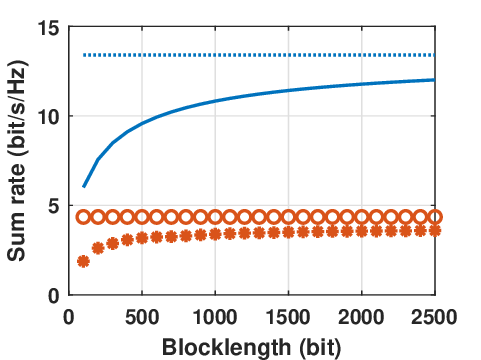}
        \caption{$\theta_1=0$}
    \end{subfigure}
    \begin{subfigure}[b]{0.48\linewidth}
        \hspace*{-0.2cm}\includegraphics[scale=0.6]{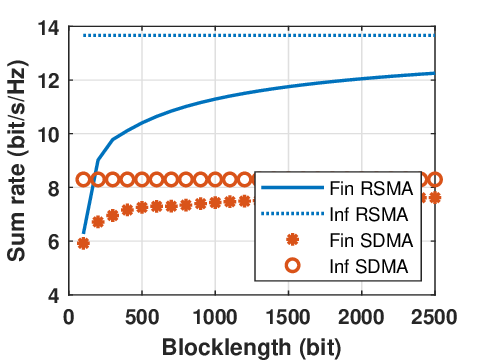}
        \caption{$\theta_1=\frac{2\pi}{18}$}
    \end{subfigure}
    \caption{Sum rate versus blocklength of different strategies for overloaded four-user deployment, $\gamma_1=0.3$.}
    \label{fig:4_user_over_bias03_angle9}
    \vspace{-5mm}
\end{figure}

 The BS is equipped with two antennas and serves four single-antenna users. The channels of users are realized as $\mathbf{h}_1={[1,1]}^H$, $\mathbf{h}_2={\gamma_1\times[1,e^{j\theta_1}]}^H$, $\mathbf{h}_3={\gamma_2\times[1,e^{j\theta_2}]}^H$, $\mathbf{h}_4={\gamma_3\times[1,e^{j\theta_3}]}^H$. $\gamma_1$, $\gamma_2$, $\gamma_3$ and $\theta_1$, $\theta_2$, $\theta_3$ are control variables. We assume that  $\gamma_1=\gamma_3$, $\gamma_2=1$, $\theta_1\in\{0,\frac{2\pi}{18}\}$, $\theta_2=\theta_1+\frac{\pi}{9}$ and $\theta_3=\theta_1+\theta_2$ for Fig. \ref{fig:4_user_over_bias03_angle9}. The individual rate constraint follows the same setting in Section \ref{Sec:Weighted sum rate}.
\par Fig. \ref{fig:4_user_over_bias03_angle9} shows that ``Fin RSMA" outperforms ``Fin SDMA" when $\gamma_1=0.3$. Besides, ``Fin RSMA" even outperforms ``Inf SDMA" since SDMA performance degrades significantly in overloaded scenarios. RSMA utilizes much smaller blocklength compared to SDMA for achieving the same sum rate. Hence, the latency is reduced greatly.
\par
\begin{figure}[t!]
    \centering
    \includegraphics[scale=0.6]{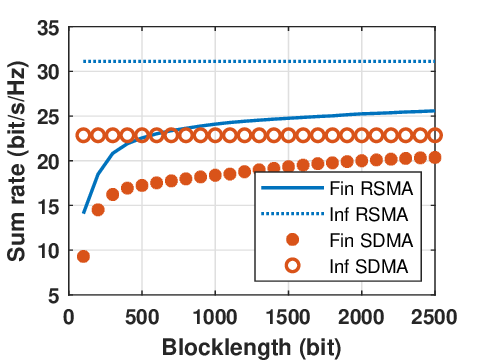}
    \caption{Sum rate versus blocklength for different strategies in overloaded eight-user deployment.}
    \label{fig:4_user_gain}
    \vspace{-7mm}
\end{figure}

 Fig. \ref{fig:4_user_gain} shows sum rate performances of RSMA and SDMA in both infinite and finite blocklength scenarios under random channel realizations. We set $N_t=4, K=8$. The channel $\mathbf{h}_k$ has independent and identically distributed (i.i.d) complex Gaussian entries with a certain variance, i.e., $\mathcal{CN}(0,\varphi_k^2)$. $\varphi_1^2=1$, $\varphi_2^2=0.875$, \ldots, $\varphi_8^2=0.125$. The sum rate performances are obtained by averaging over 100 random channel realizations. The individual rate constraint is $r_k^{th}=0.2$ bit/s/Hz for all blocklengths. The figure demonstrates that compared with finite SDMA, finite RSMA requires much smaller blocklength for achieving the same transmission rate. SDMA requires a blocklength of 2500 to achieve $20.4$ bit/s/Hz, instead RSMA only needs a blocklength of $300$ to achieve the same sum rate. RSMA can remarkably reduce blocklength (and therefore latency) in overloaded scenarios.
\vspace{-5mm}

\section{Conclusion}\label{sec:conclusion}
\vspace{-1.4mm}
This paper introduces RSMA in FBL downlink communications with low latency requirements. An optimization problem is formulated to maximize the sum rate of users. A SCA-based algorithm is adopted to solve the optimization problem. The results show that in the FBL communications, RSMA can attain the same transmission rate with smaller blocklength (and hence lower latency) in comparison with NOMA and SDMA. Particularly, in overloaded multi-antenna networks, the latency of RSMA is significantly reduced. Consequently, we conclude that RSMA is a promising strategy for low-latency communications.
\par We use the assumption of perfect CSIT to shed light on how RSMA performs with FBL. There are extensive works studying RSMA performance with infinite blocklength and imperfect CSIT. The superiority of RSMA over conventional strategies was shown in terms of CSIT inaccuracy, degree of freedom and complexity. It is expected that RSMA with FBL would have even higher performance gain compared to other schemes. The performance of FBL RSMA with imperfect CSIT is an interesting problem worth future investigation.
\bibliographystyle{IEEEtran}
\vspace{-4mm}
\bibliography{IEEEabrv,abbr_auth_reference}

\end{document}